\documentclass[twocolumn,showpacs,preprintnumbers,pra,hyperref]{revtex4}
\usepackage{amsfonts}
\usepackage{amsmath}
\usepackage{amssymb}
\usepackage{graphicx}

\begin{document}

\title{Wigner function evolution in self-Kerr Medium derived by Entangled
state representation${\small \thanks{%
Work supported by a grant from the Key Programs Foundation of Ministry of
Education of China (No. 210115) and the Research Foundation of the Education
Department of Jiangxi Province of China (No. GJJ10097).}}$}
\author{{\small Li-yun Hu\thanks{%
Corresponding author. \emph{E-mail addresses}: hlyun2008@126.com,
hlyun2008@gmail.com)}, Zheng-lu Duan, Xue-xiang Xu, and Zi-sheng Wang}}
\affiliation{{\small College of Physics \& Communication Electronics, Jiangxi Normal
University, Nanchang 330022, China}}
\date{}

\begin{abstract}
{\small By introducing the thermo entangled state representation, we convert
the calculation of Wigner function (WF) of density operator to an overlap
between "two pure" states in a two-mode enlarged Fock space. Furthermore, we
derive a new WF evolution formula of any initial state in self-Kerr Medium
with photon loss and find that the photon number distribution for any
initial state is independent of the coupling factor with Kerr Medium, where
the number state is not affected by the Kerr nonlinearity and evolves into a
density operator of binomial distribution. }

\end{abstract}

\maketitle

\section{Introduction}

Nonclassicality of optical fields has been a topic of great interest in
quantum optics and quantum information processing \cite{r1}, which is
usually associated with quantum interference and entanglement. The phase
space Wigner function (WF) \cite{1,1a} of quantum states of light is a
powerful tool for investigating such nonclassical effects. The WF was first
introduced by Wigner in 1932 to calculate quantum corrections to a classical
distribution function of a quantum-mechanical system. The partial negativity
of the WF is indeed a good indication of the highly nonclassical character
of the state \cite{2} and monitors a decoherence process of a quantum state,
e. g. the excited coherent state in both photon-loss and thermal channels
\cite{r2,r3}, the single-photon subtracted squeezed vacuum state in both
amplitude decay and phase damping channels \cite{r4}, and so on \cite%
{r5,r6,r7,r8,r8a}.

Nonlinear interaction of light in a medium provides a very useful framework
to study various nonclassical properties of quantum states of radiation. The
Kerr medium is one of the simplest nonlinearity, which shall allow us to
investigate the full time-dependent WF dynamics with or without a quantum
noise. Recently, a Fokker-Planck equation for the WF evolution in a noisy
Kerr medium ($\chi^{(3)}$ nonlinearity) is presented \cite{3}. Then the
authors numerically solved this equation assuming coherent state as an
initial condition and discussed its dissipation effects. However, for any
initial condition, as far as we are concerned, there is no report about the
WF evolution. On the other hand, by using the thermo entangled state
representation (TESR) we solved various master equations to obtain density
operators with an infinite operator-sum representation \cite{3a} and then
revealed that the WF of density operator can be expressed as an overlap
between two pure states (see Eq. (\ref{2.8}) below) \cite{4}. This brings
much convenience to calculate time evolution of WFs when quantum decoherence
happens. Thus the TESR is beneficial to quantum decoherence theory.

In this paper, we shall appeal the TESR $\left\vert \eta\right\rangle $ to
treat the WF evolution at any initial condition in self-Kerr Medium with
photon loss and present a new formula to calculate time evolution of the WF
for quantum decoherence. In addition, based on the derived WF evolution
formula, we shall deduce the photon number distribution for any initial
state in presence of Kerr interaction, where the photon number distribution
is independent of the coupling factor $\chi$ that is relative to the Kerr
medium. As examples, the WF formula is applied to the cases of initial
coherent state, and number state, respectively. Conclusions are involved in
the last section.

\section{Brief review of thermo entangled state representation}

We begin with briefly reviewing the thermo entangled state representation
(TESR). On the basis of Umezawa-Takahash thermo field dynamics (TFD)\ \cite%
{5,6,7} we constructed the TESR in the doubled Fock space \cite{7a,8,9},
\begin{align}
\left\vert \eta\right\rangle & =D\left( \eta\right) \left\vert
\eta=0\right\rangle  \notag \\
& =\exp\left[ -\frac{1}{2}|\eta|^{2}+\eta a^{\dagger}-\eta^{\ast}\tilde {a}%
^{\dagger}+a^{\dagger}\tilde{a}^{\dagger}\right] \left\vert 0,\tilde {0}%
\right\rangle ,  \label{2.1}
\end{align}
where $D\left( \eta\right) =e^{\eta a^{\dagger}-\eta^{\ast}a}$ is the
displacement operator, $\tilde{a}^{\dagger}$ is a fictitious mode
accompanying the real photon creation operator $a^{\dagger},$ $\left\vert 0,%
\tilde {0}\right\rangle =\left\vert 0\right\rangle \left\vert \tilde{0}%
\right\rangle ,$ and $\left\vert \tilde{0}\right\rangle $ is annihilated by $%
\tilde{a}$ with the relations $\left[ \tilde{a},\tilde{a}^{\dagger}\right]
=1 $ and $\left[ a,\tilde{a}^{\dagger}\right] =0$. The structure of $%
\left\vert \eta \right\rangle $ is similar to that of the EPR eigenstate\
shown in Ref. \cite{7a}. Operating $a$ and $\tilde{a}$ on $\left\vert
\eta\right\rangle $ in Eq.(\ref{2.1}) we can obtain the eigen-equations of $%
\left\vert \eta \right\rangle $,%
\begin{align}
(a-\tilde{a}^{\dagger})\left\vert \eta\right\rangle & =\eta\left\vert
\eta\right\rangle ,\;(a^{\dagger}-\tilde{a})\left\vert \eta\right\rangle
=\eta^{\ast}\left\vert \eta\right\rangle ,  \notag \\
\left\langle \eta\right\vert (a^{\dagger}-\tilde{a}) & =\eta^{\ast
}\left\langle \eta\right\vert ,\ \left\langle \eta\right\vert (a-\tilde {a}%
^{\dagger})=\eta\left\langle \eta\right\vert .  \label{2.2}
\end{align}
Note that $\left[ (a-\tilde{a}^{\dagger}),(a^{\dagger}-\tilde{a})\right] =0,$
thus $\left\vert \eta\right\rangle $ is the common eigenvector of $(a-\tilde{%
a}^{\dagger})$ and $(\tilde{a}-a^{\dagger}).$ Using the normally ordered
form of vacuum projector $\left\vert 0,\tilde{0}\right\rangle \left\langle 0,%
\tilde{0}\right\vert =\colon\exp\left( -a^{\dagger}a-\tilde {a}^{\dagger}%
\tilde{a}\right) \colon$ and the technique of integration within an ordered
product (IWOP) of operators \cite{10,11,12}, we can easily prove that $%
\left\vert \eta\right\rangle $ is complete and orthonormal,%
\begin{equation}
\int\frac{\mathtt{d}^{2}\eta}{\pi}\left\vert \eta\right\rangle \left\langle
\eta\right\vert =1,\text{ }\left\langle \eta^{\prime}\right. \left\vert
\eta\right\rangle =\pi\delta\left( \eta^{\prime}-\eta\right) \delta\left(
\eta^{\prime\ast}-\eta^{\ast}\right) .  \label{2.3}
\end{equation}
It is easily seen that $\left\vert \eta=0\right\rangle $ has the properties%
\begin{equation}
\text{ \ }\left\vert \eta=0\right\rangle =e^{a^{\dagger}\tilde{a}%
^{\dagger}}\left\vert 0,\tilde{0}\right\rangle
=\sum_{n=0}^{\infty}\left\vert n,\tilde{n}\right\rangle ,  \label{2.4}
\end{equation}
(where $n=\tilde{n}$, and $\tilde{n}$\ denotes the number in the fictitious
Hilbert space) and%
\begin{align}
a\text{\ }\left\vert \eta=0\right\rangle & =\tilde{a}^{\dagger}\left\vert
\eta=0\right\rangle ,  \notag \\
a^{\dagger}\left\vert \eta=0\right\rangle & =\tilde{a}\left\vert
\eta=0\right\rangle ,  \label{2.5} \\
\left( a^{\dagger}a\right) ^{n}\left\vert \eta=0\right\rangle & =\left(
\tilde{a}^{\dagger}\tilde{a}\right) ^{n}\left\vert \eta=0\right\rangle .
\notag
\end{align}
Note that density operators $\rho(a^{\dagger}$,$a)$ are defined in the real
space which are commutative with operators ($\tilde{a}^{\dagger}$,$\tilde{a}%
) $ in the tilde space.

In a similar way, we can introduce the state vector $\left\vert \xi
\right\rangle $ conjugated to $\left\vert \eta\right\rangle $, defined as
\begin{align}
\left\vert \xi\right\rangle & =D\left( \xi\right) e^{-a^{\dagger}\tilde {a}%
^{\dagger}}\left\vert 0,\tilde{0}\right\rangle  \notag \\
& =\exp\left( -\frac{1}{2}|\xi|^{2}+\xi a^{\dagger}+\xi^{\ast}\tilde {a}%
^{\dagger}-a^{\dagger}\tilde{a}^{\dagger}\right) \left\vert 0,\tilde {0}%
\right\rangle  \notag \\
& =(-1)^{a^{\dagger}a}\left\vert \eta=-\xi\right\rangle ,  \label{2.9}
\end{align}
which also possesses orthonormal and complete properties%
\begin{equation}
\int\frac{\mathtt{d}^{2}\xi}{\pi}\left\vert \xi\right\rangle \left\langle
\xi\right\vert =1,\text{ }\left\langle \xi^{\prime}\right. \left\vert
\xi\right\rangle =\pi\delta\left( \xi^{\prime}-\xi\right) \delta\left(
\xi^{\prime\ast}-\xi^{\ast}\right) .  \label{2.10}
\end{equation}

\section{Master equation for a self-Kerr interaction}

In the Markov approximation and interaction picture the master equation for
a dissipative cavity with Kerr medium has the form \cite{r14,r15}%
\begin{equation}
\frac{d\rho }{dt}=-i\chi \left[ \left( a^{\dagger }a\right) ^{2},\rho \right]
+\gamma \left( 2a\rho a^{\dagger }-a^{\dagger }a\rho -\rho a^{\dagger
}a\right) ,  \label{3.5}
\end{equation}%
where $\gamma $ is decaying parameter of the dissipative cavity, $\chi $ is
coupling factor depending on the Kerr medium. Milburn and Holmes \cite{15a}
solved this equation by changing it to a partial differential equation for
the Q-function\ and for an initial coherent state. Here we will solve the
master equation by virtue of the entangled state representation and present
the infinite sum representation of density operator.

Operating the both sides of Eq.(\ref{3.5}) on the state $\left\vert \eta
=0\right\rangle ,$ letting $\left\vert \rho \right\rangle =\rho \left\vert
\eta =0\right\rangle $ (Here one should understand the single-mode density
operator $\rho $\ in the left of Eq.(\ref{3.5}) as the direct product $\rho
\otimes \tilde{I}$\ when $\rho $ acts onto the two-mode state $\left\vert
\eta =0\right\rangle =e^{a^{\dagger }\tilde{a}^{\dagger }}\left\vert 0,%
\tilde{0}\right\rangle $,\ where $\tilde{I}$\ is the identity operator in
the auxiliary mode.), and using Eq.(\ref{2.5}) we can convert the master
equation in Eq. (\ref{3.5}) into the following form,

\begin{eqnarray}
\frac{d}{dt}\left\vert \rho \right\rangle &=&\left\{ -i\chi \left[ \left(
a^{\dagger }a\right) ^{2}-\left( \tilde{a}^{\dagger }\tilde{a}\right) ^{2}%
\right] \right.  \notag \\
&&+\left. \gamma \left( 2a\tilde{a}-a^{\dagger }a-\tilde{a}^{\dagger }\tilde{%
a}\right) \right\} \left\vert \rho \right\rangle ,  \label{3.6}
\end{eqnarray}%
i.e., an evolution equation of state vector $\left\vert \rho \right\rangle $%
. Its solution is then of the form%
\begin{equation}
\left\vert \rho \right\rangle =e^{-i\chi t\left[ \left( a^{\dagger }a\right)
^{2}-\left( \tilde{a}^{\dagger }\tilde{a}\right) ^{2}\right] +\gamma t\left(
2a\tilde{a}-a^{\dagger }a-\tilde{a}^{\dagger }\tilde{a}\right) }\left\vert
\rho _{0}\right\rangle ,  \label{3.7}
\end{equation}%
where $\left\vert \rho _{0}\right\rangle =\rho _{0}\left\vert \eta
=0\right\rangle ,$ $\rho _{0}$ is an initial density operator. The advantage
of using thermo field notation over more traditional algebraic manipulation
with superoperators is that in many situations (and, particularly, ones of
our interest) it enables to simplify, make more illustrative and less
cumbersome finding the solution (\ref{3.7}) and estimation of time-dependent
matrix elements. In particular, it allows to represent in a simple form a
factorization of the superoperator exp\{$\cdots $\} into multipliers with
easily estimated actions on the number states \cite{15b}.

By introducing the following operators,
\begin{equation}
K_{0}=a^{\dagger }a-\tilde{a}^{\dagger }\tilde{a},\text{\ }K_{z}=\frac{%
a^{\dagger }a+\tilde{a}^{\dagger }\tilde{a}+1}{2},\text{\ }K_{-}=a\tilde{a},
\label{3.8}
\end{equation}%
which satisfy $\left[ K_{0},K_{z}\right] =\left[ K_{0},K_{-}\right] =0,$ we
can rewrite Eq.(\ref{3.7}) as%
\begin{align}
\left\vert \rho \right\rangle & =e^{\left\{ -i\chi t\left[ K_{0}(2K_{z}-1)%
\right] +\gamma t\left( 2K_{-}-2K_{z}+1\right) \right\} }\left\vert \rho
_{0}\right\rangle  \notag \\
& =\exp \left[ i\chi tK_{0}+\gamma t\right]  \notag \\
& \times \exp \left\{ -2t\left( \gamma +i\chi K_{0}\right) \left[ K_{z}+%
\frac{-\gamma }{\gamma +i\chi K_{0}}K_{-}\right] \right\} \left\vert \rho
_{0}\right\rangle .  \label{3.9}
\end{align}%
With the aid of the operator identity \cite{r16}%
\begin{eqnarray}
e^{\lambda \left( A+\sigma B\right) } &=&e^{\lambda A}\exp \left[ \sigma
B\left( 1-e^{-\lambda \tau }\right) /\tau \right]  \notag \\
&=&\exp \left[ \sigma B\left( e^{\lambda \tau }-1\right) /\tau \right]
e^{\lambda A},  \label{3.10}
\end{eqnarray}%
which is valid for $\left[ A,B\right] =\tau B,$ and noticing $\left[
K_{z},K_{-}\right] =-K_{-},$\ we can reform Eq.(\ref{3.9}) as%
\begin{equation}
\left\vert \rho \right\rangle =\exp \left[ i\chi tK_{0}+\gamma t\right] \exp %
\left[ \Gamma _{z}K_{z}\right] \exp \left[ \Gamma _{-}K_{-}\right]
\left\vert \rho _{0}\right\rangle ,  \label{2.12}
\end{equation}%
where%
\begin{equation}
\Gamma _{z}=-2t\left( \gamma +i\chi K_{0}\right) ,\text{ }\Gamma _{-}=\frac{%
\gamma (1-e^{-2t\left( \gamma +i\chi K_{0}\right) })}{\gamma +i\chi K_{0}}.
\label{2.13}
\end{equation}%
From Eq.(\ref{2.12}) we can obtain the infinite operator-sum form of $\rho
\left( t\right) $, (see Appendix A)
\begin{equation}
\rho \left( t\right) =\sum_{m,n,l=0}^{\infty }M_{m,n,l}\rho _{0}\mathcal{M}%
_{m,n,l}^{\dagger },  \label{2.26}
\end{equation}%
where the two operators $M_{m,n,l}$ and $\mathcal{M}_{m,n,l}^{\dagger }$ are
respectively defined as%
\begin{align}
M_{m,n,l}& \equiv \sqrt{\frac{\Lambda _{m,n}^{l}}{l!}}e^{-i\chi
tm^{2}-\gamma tm}\left\vert m\right\rangle \left\langle m\right\vert a^{l},%
\text{ }  \notag \\
\mathcal{M}_{m,n,l}^{\dagger }& \equiv \left\{ \sqrt{\frac{\Lambda _{n,m}^{l}%
}{l!}}e^{-i\chi tn^{2}-\gamma tn}\left\vert n\right\rangle \left\langle
n\right\vert a^{l}\right\} ^{\dag }.  \label{2.27}
\end{align}%
Although $M_{m,n,l}$ is not hermite conjugate to $\mathcal{M}%
_{m,n,l}^{\dagger }$, the normalization still holds, ($\sum_{m,n,l=0}^{%
\infty }\mathcal{M}_{m,n,l}^{\dagger }M_{m,n,l}=1,$ see Appendix B \cite%
{ijtp}) i.e., they are trace-preserving in a general sense, so $M_{m,n,l}$
and $\mathcal{M}_{m,n,l}^{\dagger }$ may be named the generalized Kraus
operators.

\section{Evolution of Wigner function for self-Kerr channel}

In this section, we consider Wigner function's time evolution in the
self-Kerr medium channel. For this purpose, we shall derive a new expression
of Wigner function in the TESR. According to the definition of Wigner
function of density operator $\rho,$
\begin{equation}
W\left( \alpha,\alpha^{\ast}\right) =\text{Tr}\left[ \Delta\left(
\alpha,\alpha^{\ast}\right) \rho\right] ,  \label{2.6}
\end{equation}
where $\Delta\left( \alpha,\alpha^{\ast}\right) $ is the single-mode Wigner
operator \cite{1,r16}, whose explicit normally ordered form is \cite{14b}
\begin{equation}
\Delta\left( \alpha,\alpha^{\ast}\right) =\frac{1}{\pi}\colon e^{-2\left(
a^{\dagger}-\alpha^{\ast}\right) \left( a-\alpha\right) }\colon=\frac {1}{\pi%
}D\left( 2\alpha\right) (-1)^{a^{\dagger}a}.  \label{2.7}
\end{equation}
By using $\left\langle \tilde{n}\right\vert \left. \tilde{m}\right\rangle
=\delta_{n,m}$ ($n=\tilde{n},m=\tilde{m}$) and noticing (\ref{2.9}) as well
as $\left\vert \rho\right\rangle =\rho\left\vert \eta=0\right\rangle ,$%
\textbf{\ }we can reform Eq.(\ref{2.6}) as%
\begin{align}
W\left( \alpha,\alpha^{\ast}\right) & =\sum_{m,n}^{\infty}\left\langle n,%
\tilde{n}\right\vert \Delta\left( \alpha,\alpha^{\ast}\right) \rho\left\vert
m,\tilde{m}\right\rangle  \notag \\
& =\frac{1}{\pi}\left\langle \eta=0\right\vert D\left( 2\alpha\right)
(-1)^{a^{\dagger}a}\left\vert \rho\right\rangle  \notag \\
& =\frac{1}{\pi}\left\langle \eta=-2\alpha\right\vert
(-1)^{a^{\dagger}a}\left\vert \rho\right\rangle  \notag \\
& =\frac{1}{\pi}\left\langle \xi=2\alpha\right\vert \left. \rho\right\rangle
,  \label{2.8}
\end{align}
where Eq. (\ref{2.8}) is the Wigner function formula in thermo entangled
state representation, with which the Wigner function of density operator is
simplified as an overlap between two \textquotedblleft pure
states\textquotedblright\ in enlarged Fock space, rather than using ensemble
average in the system-mode space. This will brings much convenience to
calculate the time evolution of Wigner functions when quantum decoherence
happens.

Projecting (\ref{2.12}) on the entangled state representation $\frac{1}{\pi }%
\left\langle \xi _{=2\alpha }\right\vert ,$ and inserting the completeness
relation (\ref{2.10}), we find
\begin{equation}
W\left( \alpha ,\alpha ^{\ast },t\right) =4\int \frac{\mathtt{d}^{2}\beta }{%
\pi }G\left( \alpha ,\beta ,t\right) W\left( \beta ,\beta ^{\ast },0\right) ,
\label{3.1}
\end{equation}%
where $W\left( \alpha ,\alpha ^{\ast },t\right) $ and $W\left( \beta ,\beta
^{\ast },0\right) $ are the Wigner functions at the evolving time $t$ and
initial time, respectively, and
\begin{eqnarray}
G\left( \alpha ,\beta ,t\right) &=&\left\langle \xi _{=2\alpha }\right\vert
\exp \left[ i\chi tK_{0}+\gamma t\right]  \notag \\
&&\times \exp \left[ \Gamma _{z}K_{z}\right] \exp \left[ \Gamma _{-}K_{-}%
\right] \left\vert \xi _{=2\beta }^{\prime }\right\rangle .  \label{3.2}
\end{eqnarray}%
It is convenient to the matrix element in (\ref{3.2}) according to the
two-mode Fock space. Thus the $\left\langle \xi _{=2\alpha }\right\vert $ is
expanded as
\begin{equation}
\left\langle \xi \right\vert =\left\langle 0,\tilde{0}\right\vert
\sum_{m,n=0}^{\infty }\frac{a^{m}\tilde{a}^{n}}{m!n!}H_{m,n}\left( \xi
^{\ast },\xi \right) e^{-\left\vert \xi \right\vert ^{2}/2}.  \label{3.3}
\end{equation}%
By using the two-mode Fock state $\left\vert m,\tilde{n}\right\rangle
=a^{\dag m}\tilde{a}^{\dag n}/\sqrt{m!n!}\left\vert 0,\tilde{0}\right\rangle
$, we get
\begin{equation}
\left\langle \xi \right\vert \left. m,\tilde{n}\right\rangle =H_{m,n}\left(
\xi ^{\ast },\xi \right) e^{-\left\vert \xi \right\vert ^{2}/2}/\sqrt{m!n!},
\label{3.4}
\end{equation}%
where $H_{m,n}\left( \xi ^{\ast },\xi \right) $ is the two-variable Hermite
polynomials \cite{26,27}. Inserting the complete relation $%
\sum_{m,n=0}^{\infty }\left\vert m,\tilde{n}\right\rangle \left\langle m,%
\tilde{n}\right\vert =1,$ after a long but straight calculation, then the
Wigner function's evolution is given by (see Appendix C)%
\begin{equation}
W\left( \alpha ,\alpha ^{\ast },t\right) =\sum_{m,n=0}^{\infty
}C_{m,n}\left( \alpha ,\alpha ^{\ast },t\right) E_{m,n},  \label{3.11}
\end{equation}%
where $\Lambda _{m,n}\equiv \frac{\gamma (1-e^{-2t\left( \gamma +i\chi
\left( m-n\right) \right) })}{\gamma +i\chi \left( m-n\right) },$%
\begin{eqnarray}
&&C_{m,n}\left( \alpha ,\alpha ^{\ast },t\right)  \notag \\
&\equiv &\frac{e^{-i\chi t\left( m^{2}-n^{2}\right) -\gamma t\left(
m+n\right) }e^{-2\left\vert \alpha \right\vert ^{2}}}{m!n!\left( \Lambda
_{m,n}+1\right) ^{(m+n+2)/2}}H_{m,n}\left( 2\alpha ^{\ast },2\alpha \right) ,
\label{3.12}
\end{eqnarray}%
and%
\begin{eqnarray}
E_{m,n} &=&4\int \frac{\mathtt{d}^{2}\beta }{\pi }W\left( \beta ,\beta
^{\ast },0\right) e^{\frac{2\left( \Lambda _{m,n}-1\right) }{\Lambda _{m,n}+1%
}\left\vert \beta \right\vert ^{2}}  \notag \\
&&\times H_{m,n}\left( \frac{2\beta }{\sqrt{\Lambda _{m,n}+1}},\frac{2\beta
^{\ast }}{\sqrt{\Lambda _{m,n}+1}}\right) .  \label{3.13}
\end{eqnarray}%
It is obvious that, when $\chi =0,$ the case of photon loss, $\Lambda
_{m,n}\rightarrow (1-e^{-2\gamma t})=T$ and Eq.(\ref{3.11}) just does reduce
to (see Appendix E)
\begin{equation}
W\left( \alpha ,\alpha ^{\ast },t\right) =\frac{2}{T}\int \frac{\mathtt{d}%
^{2}\beta }{\pi }e^{-\allowbreak \frac{2}{T}\left\vert \alpha -\beta
e^{-\gamma t}\right\vert ^{2}}W\left( \beta ,\beta ^{\ast },0\right) ,
\label{3.14}
\end{equation}%
which is just the evolving formula of Wigner function for amplitude-damping
channel. While for $\gamma =0,$ without photon-loss, Eq.(\ref{3.11}) reduces
to
\begin{align}
& W\left( \alpha ,\alpha ^{\ast },t\right)  \notag \\
& =\sum_{m,n=0}^{\infty }\frac{\exp \left[ -i\chi t\left( m^{2}-n^{2}\right) %
\right] }{m!n!e^{2\left\vert \alpha \right\vert ^{2}}}H_{m,n}\left( 2\alpha
^{\ast },2\alpha \right)  \notag \\
& \times 4\int \frac{\mathtt{d}^{2}\beta }{\pi }e^{-2\left\vert \beta
\right\vert ^{2}}H_{m,n}\left( 2\beta ,2\beta ^{\ast }\right) W\left( \beta
,\beta ^{\ast },0\right) .  \label{3.15}
\end{align}

\section{Photon number distribution in presence of Kerr interaction}

Now we consider photon number (PN) distribution in presence of Kerr medium.
According to the TFD, we can reform the PN $p\left( n\right) =\mathtt{tr}%
\left[ \rho \left\vert n\right\rangle \left\langle n\right\vert \right] $ as
\begin{eqnarray}
p\left( n\right) &=&\left\langle n\right\vert \rho \left\vert n\right\rangle
=\sum_{m=0}^{\infty }\left\langle n,\tilde{n}\right\vert \rho \left\vert m,%
\tilde{m}\right\rangle  \notag \\
&=&\left\langle n,\tilde{n}\right\vert \rho \left\vert \eta =0\right\rangle
=\left\langle n,\tilde{n}\right\vert \left. \rho \right\rangle ,  \label{5.1}
\end{eqnarray}%
which is converted to the matrix element $\left\langle n,\tilde{n}%
\right\vert \left. \rho \right\rangle $ in the context of thermo dynamics.
Then using the completeness of $\left\langle \xi \right\vert $ and Eq.(\ref%
{5.1}) as well as Eq.(\ref{2.8}), we have%
\begin{align}
p\left( n\right) & =\int \frac{\mathtt{d}^{2}\xi }{\pi }\left\langle n,%
\tilde{n}\right\vert \left. \xi \right\rangle \left\langle \xi \right\vert
\left. \rho \right\rangle  \notag \\
& =\int \mathtt{d}^{2}\xi \left\langle n,\tilde{n}\right\vert \left. \xi
\right\rangle W\left( \alpha =\xi /2,\alpha ^{\ast }=\xi ^{\ast }/2\right)
\notag \\
& =4\pi \int \mathtt{d}^{2}\alpha W_{\left\vert n\right\rangle \left\langle
n\right\vert }\left( \alpha ,\alpha ^{\ast }\right) W\left( \alpha ,\alpha
^{\ast }\right) ,  \label{5.2}
\end{align}%
where $W_{\left\vert n\right\rangle \left\langle n\right\vert }\left( \alpha
,\alpha ^{\ast }\right) =\frac{(-1)^{n}}{\pi }e^{-2\left\vert \alpha
\right\vert ^{2}}L_{n}(4\left\vert \alpha \right\vert ^{2})$ is the Wigner
function of number state $\left\vert n\right\rangle \left\langle
n\right\vert $ as shown in \cite{28,29}. Thus one can calculate the PN by
combining Eqs.(\ref{3.11}) and (\ref{5.2}).

Next we evaluate the PN for the above decoherence model in Eq.(\ref{3.5}).
Substituting Eq.(\ref{3.11}) into Eq.(\ref{5.2}), we have
\begin{align}
p\left( s\right) & =4\pi \sum_{m,n=0}^{\infty }\int \mathtt{d}^{2}\alpha
W_{\left\vert s\right\rangle \left\langle s\right\vert }\left( \alpha
,\alpha ^{\ast }\right) C_{m,n}\left( \alpha ,\alpha ^{\ast },t\right)
E_{m,n}  \notag \\
& =\sum_{m,n=0}^{\infty }\frac{4\pi e^{-i\chi t\left( m^{2}-n^{2}\right)
-\gamma t\left( m+n\right) }}{m!n!\left( \Lambda _{m,n}+1\right) ^{(m+n+2)/2}%
}E_{m,n}\cdot F_{m,n},  \label{5.3}
\end{align}%
where
\begin{eqnarray}
F_{m,n} &\equiv &\int \mathtt{d}^{2}\alpha e^{-2\left\vert \alpha
\right\vert ^{2}}W_{\left\vert s\right\rangle \left\langle s\right\vert
}\left( \alpha ,\alpha ^{\ast }\right) H_{m,n}\left( 2\alpha ^{\ast
},2\alpha \right)  \notag \\
&=&\frac{s!}{4}\delta _{m,s}\delta _{n,s}.  \label{5.4}
\end{eqnarray}%
Then substituting Eqs.(\ref{5.4}) and (\ref{3.13}) into (\ref{5.3}) yields
\begin{eqnarray}
p\left( s\right) &=&\frac{4(-1)^{s}e^{2\gamma t}}{\left( 2e^{2\gamma
t}-1\right) ^{s+1}}\int \mathtt{d}^{2}\beta \exp \left\{ -\frac{2\left\vert
\beta \right\vert ^{2}}{2e^{2\gamma t}-1}\right\}  \notag \\
&&\text{ \ \ \ \ \ \ \ \ \ }\times L_{s}\left( \frac{4e^{2\gamma
t}\left\vert \beta \right\vert ^{2}}{2e^{2\gamma t}-1}\right) W\left( \beta
,\beta ^{\ast },0\right) ,  \label{5.5}
\end{eqnarray}%
which corresponds to the photon number of density operator in the
amplitude-damping quantum channel \cite{4}. From Eq.(\ref{5.5}), it is
easily to see that, for any initial state, the photon number distribution $%
p\left( s\right) $ is independent of the coupling factor $\chi $ that is
relative to the Kerr medium, as respected in \cite{30}.

\section{Evolution of quantum states}

The phase space Wigner distribution function description of quantum states
of light is a powerful tool to investigate nonclassical effects, such as
quantum interference and entanglement. In this section, as the applications
of WF evolution formula, we take two special initial states as examples.

(1) When the initial state is the coherent state $\left\vert z\right\rangle $%
, whose WF is given $W\left( \beta ,\beta ^{\ast }\right) =\frac{1}{\pi }%
e^{-2\left\vert \beta -z\right\vert ^{2}},$ thus substituting into Eq.(\ref%
{3.13}) yields (see Appendix G)%
\begin{equation}
E_{m,n}=\frac{1}{\pi }\left( \Lambda _{m,n}+1\right) ^{\frac{m+n+2}{2}%
}e^{\left( \allowbreak \Lambda _{m,n}-1\right) \left\vert z\right\vert
^{2}}z^{m}z^{\ast n},  \label{5.6}
\end{equation}%
so%
\begin{eqnarray}
&&W\left( \alpha ,\alpha ^{\ast },t\right)  \notag \\
&=&\frac{e^{-2\left\vert \alpha \right\vert ^{2}}}{\pi }\sum_{m,n=0}^{\infty
}\frac{z^{m}z^{\ast n}}{m!n!}e^{-i\chi t\left( m^{2}-n^{2}\right) -\gamma
t\left( m+n\right) }  \notag \\
&&\times e^{\left( \allowbreak \Lambda _{m,n}-1\right) \left\vert
z\right\vert ^{2}}H_{m,n}\left( 2\alpha ^{\ast },2\alpha \right) ,
\label{5.7}
\end{eqnarray}%
which is a new expression of the evolution of WF for any initial state. In
particular, when $\gamma =0,$ without the dissipation, Eq.(\ref{5.7})
reduces to
\begin{eqnarray}
&&W\left( \alpha ,\alpha ^{\ast },t\right)  \notag \\
&=&\frac{e^{-\left\vert z\right\vert ^{2}}}{\pi e^{2\left\vert \alpha
\right\vert ^{2}}}\sum_{m,n=0}^{\infty }\frac{z^{m}z^{\ast n}}{m!n!}%
e^{-i\chi t(m^{2}-n^{2})}H_{m,n}\left( 2\alpha ^{\ast },2\alpha \right) ,
\label{5.8}
\end{eqnarray}%
which is identical to Eq.(7) in Ref.\cite{3}, where Eq.(7) is used to make
the numerical calculation since it is much more rapid than the other
expression Eq.(6). In addition, from Eq.(\ref{5.7}) one can see that the WF
can be obtained very quickly when the dissipation cannot be negligible.
Further when $\chi t=2\pi $, Eq.(\ref{5.8}) just returns to the WF of the
initial coherent state.

Fig.1 presents the plots of the WF for different parameters $\chi t$ and $%
\alpha=2\mathtt{.}$ From Fig.1, one can see that the WF turns into an
ellipse and squeezing appears in an appropriate direction. Then the ellipse
changes into a banana shape and a tailor of the interference fringes appears
where the distribution takes the negative values.

\begin{figure}[tbp]
\label{Fig8.1} \centering
\includegraphics[width=3.0in]{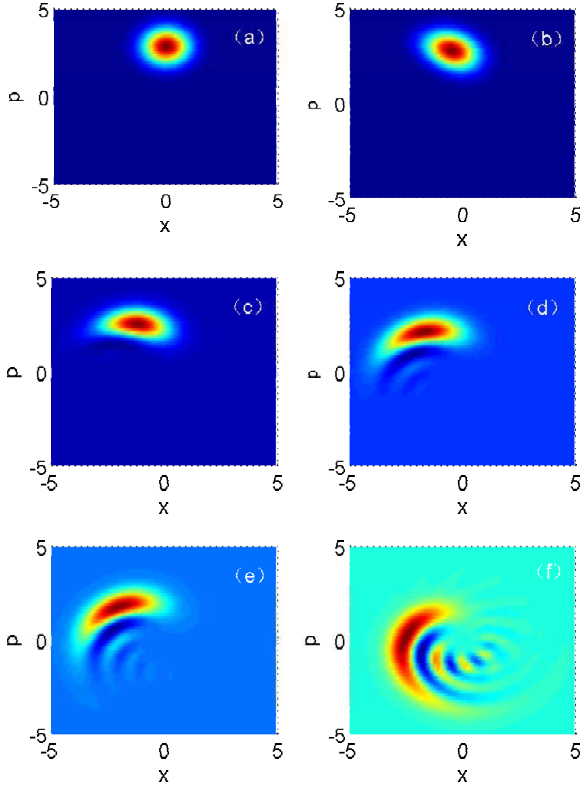}
\caption{(Color online) The WF for different parameters $\protect\chi t$ and
$\protect\alpha =2\mathtt{,(a)}$ $\protect\chi t=0,\mathtt{(b)}$ $\protect%
\chi t=0.04,\mathtt{(c)}$ $\protect\chi t=0.06,\mathtt{(d)}$ $\protect\chi %
t=0.08,\mathtt{(e)}$ $\protect\chi t=0.1,\mathtt{(f)}$ $\protect\chi t=0.2.$}
\end{figure}

(2) Another example is number state, where the WF\quad of number state $%
\left\vert s\right\rangle $ is given by \
\begin{eqnarray}
W_{s}\left( \beta ,\beta ^{\ast },0\right) &=&\frac{1/s!}{\pi }%
e^{-2\left\vert \beta \right\vert ^{2}}H_{s,s}\left( 2\beta ,2\beta ^{\ast
}\right)  \notag \\
&=&\frac{(-1)^{s}}{\pi }e^{-2\left\vert \beta \right\vert
^{2}}L_{s}(4\left\vert \beta \right\vert ^{2}),  \label{5.9}
\end{eqnarray}%
substituting it into (\ref{3.13}) and using the generating function of
two-variable Hermite polynomials [see below Eq.(F3)], we have%
\begin{equation}
E_{m,n}=\frac{s!}{\pi }\frac{\Lambda _{m,n}^{s-m}\left( \Lambda
_{m,n}+1\right) ^{m+1}}{\left( s-m\right) !}\delta _{m,n},  \label{5.10}
\end{equation}%
thus the evolution of WF for $\left\vert s\right\rangle $ is%
\begin{eqnarray}
&&W_{s}\left( \alpha ,\alpha ^{\ast },t\right)  \notag \\
&=&\sum_{m=0}^{s}\binom{s}{m}e^{-2m\gamma t}\left( 1-e^{-2\gamma t}\right)
^{s-m}W_{m}\left( \alpha ,\alpha ^{\ast },0\right) ,  \label{5.11}
\end{eqnarray}%
which is the WF of number state in the photon-loss channel and indicates
that the number state is not affected by the Kerr nonlinearity. In
particular, when $\gamma =0,$ or $t=0,$ Eq.(\ref{5.11}) just reduces to the
WF of number state. From Eq.(\ref{5.11}), on the other hand, it is found
that the number state evolves into a density operator of binomial
distribution (a mixed state) if $e^{-2\gamma t}$ is a binomial parameter.

\section{Conclusions}

In summary, by converting Wigner function for quantum state into an overlap
between two "pure states" in a two-mode enlarged Fock space, we investigate
the WF evolution of any initial condition in self-Kerr Medium with photon
loss and present a new formula for calculating time evolution of the WF for
quantum decoherence. Based on the derived WF evolution formula, in addition,
we discuss the photon number distribution for any initial state in presence
of Kerr interaction. It is found that the photon number distribution is
independent of the coupling factor $\chi $ in correlation with the Kerr
medium, as expected by people. As applications, furthermore, the two cases
of initial coherent state and number state are considered. It is shown that
the coherent state can be squeezed due to the presence of Kerr medium, while
the number state is not affected by the Kerr nonlinearity and evolves into a
density operator of binomial distribution (a mixed state) with $e^{-2\gamma
t}$ being a binomial parameter.

\bigskip

\textbf{ACKNOWLEDGEMENTS} Work supported by a grant from the Key Programs
Foundation of Ministry of Education of China (No. 210115) and the Research
Foundation of the Education Department of Jiangxi Province of China (No.
GJJ10097).

\bigskip

\textbf{Appendix A: Derivation of Eq.(\ref{2.26})}

In order to obtain the infinite operator-sum form of $\rho \left( t\right) $
from Eq.(\ref{2.12}), using the completeness relation of Fock state in the
enlarged space $\sum_{m,n=0}^{\infty }\left\vert m,\tilde{n}\right\rangle
\left\langle m,\tilde{n}\right\vert =1$ and noticing $a^{\dagger
l}\left\vert n\right\rangle =\sqrt{\frac{\left( l+n\right) !}{n!}}\left\vert
n+l\right\rangle $, we have%
\begin{align}
\left\vert \rho \right\rangle & =e^{i\chi tK_{0}+\gamma t}e^{\Gamma
_{z}K_{z}}e^{\Gamma _{-}K_{-}}\left\vert \rho _{0}\right\rangle   \notag \\
& =e^{i\chi tK_{0}+\gamma t}e^{\Gamma _{z}K_{z}}\sum_{m,n=0}^{\infty
}\left\vert m,\tilde{n}\right\rangle \left\langle m,\tilde{n}\right\vert
e^{\Gamma _{-}K_{-}}\left\vert \rho _{0}\right\rangle   \notag \\
& =\sum_{m,n=0}^{\infty }e^{-i\chi t\left( m^{2}-n^{2}\right) -\gamma
t\left( m+n\right) }  \notag \\
& \times \left\vert m,\tilde{n}\right\rangle \left\langle m,\tilde{n}%
\right\vert e^{\Lambda _{m,n}a\tilde{a}}\left\vert \rho _{0}\right\rangle ,
\tag{A1}
\end{align}%
where
\begin{equation}
\Lambda _{m,n}=\frac{\gamma (1-e^{-2t\left( \gamma +i\chi \left( m-n\right)
\right) })}{\gamma +i\chi \left( m-n\right) }.  \tag{A2}
\end{equation}%
Furthermore, using the relations
\begin{equation}
\left\langle n\right\vert \left. \eta =0\right\rangle =\left\vert \tilde{n}%
\right\rangle ,\text{ }\left\vert m,\tilde{n}\right\rangle =\left\vert
m\right\rangle \left\langle n\right\vert \left. \eta =0\right\rangle ,
\tag{A3}
\end{equation}%
we find
\begin{align}
\left\langle m,\tilde{n}\right\vert a^{l}\rho _{0}a^{\dagger l}\left\vert
\eta =0\right\rangle & =\left\langle m\right\vert \left\langle \tilde{n}%
\right\vert a^{l}\rho _{0}a^{\dagger l}\left\vert \eta =0\right\rangle
\notag \\
& =\left\langle m\right\vert a^{l}\rho _{0}a^{\dagger l}\left( \left\langle
\tilde{n}\right\vert \left. \eta =0\right\rangle \right)   \notag \\
& =\left\langle m\right\vert a^{l}\rho _{0}a^{\dagger l}\left\vert
n\right\rangle .  \tag{A4}
\end{align}%
Thus Eq.(A1) becomes
\begin{align}
\left\vert \rho \right\rangle & =\sum_{m,n,l=0}^{\infty }\frac{\Lambda
_{m,n}^{l}}{l!}e^{-i\chi t\left( m^{2}-n^{2}\right) -\gamma t\left(
m+n\right) }  \notag \\
& \times \left\vert m,\tilde{n}\right\rangle \left\langle m\right\vert
a^{l}\rho _{0}a^{\dagger l}\left\vert n\right\rangle   \notag \\
& =\sum_{m,n,l=0}^{\infty }\frac{\sqrt{\left( n+l\right) !\left( m+l\right) !%
}}{\sqrt{n!m!}l!\Lambda _{m,n}^{-l}}  \notag \\
& \times e^{-i\chi t\left( m^{2}-n^{2}\right) -\gamma t\left( m+n\right)
}\left\vert m,\tilde{n}\right\rangle \rho _{0,m+l,n+l},  \tag{A5}
\end{align}%
where $\rho _{0,m+l,n+l}\equiv \left\langle m+l\right\vert \rho
_{0}\left\vert n+l\right\rangle .$ Using Eq.(A3) again, we see%
\begin{align}
\left\vert \rho \right\rangle & =\sum_{m,n,l=0}^{\infty }\frac{\sqrt{\left(
n+l\right) !\left( m+l\right) !}}{\sqrt{n!m!}l!\Lambda _{m,n}^{-l}}  \notag
\\
& \times e^{-i\chi t\left( m^{2}-n^{2}\right) -\gamma t\left( m+n\right)
}\rho _{0,m+l,n+l}\left\vert m\right\rangle \left\langle n\right\vert \left.
\eta =0\right\rangle .  \tag{A6}
\end{align}%
After depriving $\left\vert \eta =0\right\rangle $ from the both sides of
Eq.(A6), the solution of master equation (\ref{3.5}) appears as an infinite
operator-sum form%
\begin{widetext}
\begin{align}
\rho \left( t\right) & =\sum_{m,n,l=0}^{\infty }\sqrt{\frac{\left(
n+l\right) !\left( m+l\right) !}{n!m!}}\frac{\Lambda _{m,n}^{l}}{l!}%
e^{-i\chi t\left( m^{2}-n^{2}\right) -\gamma t\left( m+n\right) }\left\vert
m\right\rangle \left\langle m+l\right\vert \rho _{0}\left\vert
n+l\right\rangle \left\langle n\right\vert  \notag \\
& =\sum_{m,n,l=0}^{\infty }\frac{\Lambda _{m,n}^{l}}{l!}e^{-i\chi t\left(
m^{2}-n^{2}\right) -\gamma t\left( m+n\right) }\left\vert m\right\rangle
\left\langle m\right\vert a^{l}\rho _{0}a^{\dagger l}\left\vert
n\right\rangle \left\langle n\right\vert .  \tag{A7}
\end{align}%
\end{widetext}Note that the factor $\left( m-n\right) $ appears in the
denominator of $\Lambda _{m,n}$ (see Eq.(A2)), (this is originated from the
nonlinear term of $\left( a^{\dagger }a\right) ^{2}$) so that it is
impossible to move all $n-$dependent$\ $terms to the right of $a^{l}\rho
_{0}a^{\dagger l}$. Fortunately, we can formally express Eq.(A7) as Eq.(\ref%
{2.26}).

\textbf{Appendix B: Proof of normalization for the generalized Kraus
operators}

Using the operator identity $e^{\lambda a^{\dagger}a}=\colon\exp\left[
\left( e^{\lambda}-1\right) a^{\dagger}a\right] \colon$ and the IWOP
technique, we can prove that
\begin{align}
& \ \sum_{m,n,l=0}^{\infty}\mathcal{M}_{m,n,l}^{\dagger}M_{m,n,l}  \notag \\
& =\sum_{n,l=0}^{\infty}\frac{\left( n+l\right) !}{n!}\frac{(1-e^{-2t\gamma
})^{l}}{l!}e^{-2n\gamma t}\left\vert n+l\right\rangle \left\langle
n+l\right\vert  \notag \\
& =\sum_{n,l=0}^{\infty}\frac{(1-e^{-2t\gamma})^{l}}{l!}a^{\dag
l}e^{-2\gamma ta^{\dagger}a}\left\vert n\right\rangle \left\langle
n\right\vert a^{l}  \notag \\
& =\sum_{l=0}^{\infty}\frac{(1-e^{-2t\gamma})^{l}}{l!}\colon\exp\left[
\left( e^{-2\gamma t}-1\right) a^{\dagger}a\right] \left( a^{\dagger
}a\right) ^{l}\colon=1,  \tag{B1}
\end{align}
from which one can see that the normalization still holds, i.e., they are
trace-preserving in a general sense, so $M_{m,n,l}$ and $\mathcal{M}%
_{m,n,l}^{\dagger}$ may be named the generalized Kraus operators.

\textbf{Appendix C: Derivation of Eq.(\ref{3.11})}

Using Eq.(\ref{3.4}) and (A1) as well as (A2), Eq.(\ref{3.2}) can be
rewritten as
\begin{widetext}
\begin{align}
G\left( \alpha ,\beta ,t\right) & =\sum_{m,n=0}^{\infty }e^{-i\chi t\left(
m^{2}-n^{2}\right) -\gamma t\left( m+n\right) }\left\langle \xi _{=2\alpha
}\right. \left\vert m,\tilde{n}\right\rangle \left\langle m,\tilde{n}%
\right\vert e^{\Lambda _{m,n}a\tilde{a}}\left\vert \xi _{=2\beta }^{\prime
}\right\rangle  \notag \\
& =\sum_{m,n=0}^{\infty }\frac{e^{-2\left\vert \alpha \right\vert ^{2}}}{%
\sqrt{m!n!}}e^{-i\chi t\left( m^{2}-n^{2}\right) -\gamma t\left( m+n\right)
}H_{m,n}\left( 2\alpha ^{\ast },2\alpha \right) \sum_{l=0}^{\infty }\frac{%
\Lambda _{m,n}^{l}}{l!}\left\langle m,\tilde{n}\right\vert a^{l}\tilde{a}%
^{l}\left\vert \xi _{=2\beta }^{\prime }\right\rangle  \notag \\
& =e^{-2\left\vert \beta \right\vert ^{2}-2\left\vert \alpha \right\vert
^{2}}\sum_{m,n=0}^{\infty }\frac{e^{-i\chi t\left( m^{2}-n^{2}\right)
-\gamma t\left( m+n\right) }}{m!n!}H_{m,n}\left( 2\alpha ^{\ast },2\alpha
\right) \sum_{l=0}^{\infty }\frac{\Lambda _{m,n}^{l}}{l!}H_{m+l,n+l}\left(
2\beta ,2\beta ^{\ast }\right) .  \tag{C1}
\end{align}
\end{widetext}
Further using a new sum formula (see appendix D)
\begin{align}
& \sum_{l=0}^{\infty }\frac{z^{l}}{l!}H_{m+l,n+l}\left( x,y\right)  \notag \\
& =\frac{e^{\frac{z\allowbreak xy}{z+1}}}{\left( z+1\right) ^{(m+n+2)/2}}%
H_{m,n}\left( \frac{x}{\sqrt{z+1}},\frac{y}{\sqrt{z+1}}\right) ,  \tag{C2}
\end{align}%
thus Eq.(C1) can be recast into the following form%
\begin{align}
G\left( \alpha ,\beta ,t\right) & =\sum_{m,n=0}^{\infty }C_{m,n}\left(
\alpha ,\alpha ^{\ast },t\right) e^{2\frac{\Lambda _{m,n}-1}{\Lambda _{m,n}+1%
}\left\vert \beta \right\vert ^{2}}  \notag \\
& \times H_{m,n}\left( \frac{2\beta }{\sqrt{\Lambda _{m,n}+1}},\frac{2\beta
^{\ast }}{\sqrt{\Lambda _{m,n}+1}}\right) ,  \tag{C3}
\end{align}%
where $C_{m,n}\left( \alpha ,\alpha ^{\ast },t\right) $ is defined in (\ref%
{3.12}). Substituting Eq.(C3) into (\ref{3.1}) yields (\ref{3.11}) and (\ref%
{3.13}).

\textbf{Appendix D: Derivation of Eq.(C2)}

Using the integratal expression of two-mode Hermite polynomials,
\begin{equation}
H_{m,n}\left( \xi ,\eta \right) =(-1)^{n}e^{\xi \eta }\int \frac{d^{2}z}{\pi
}z^{n}z^{\ast m}e^{-\left\vert z\right\vert ^{2}+\xi z-\eta z^{\ast }},
\tag{D1}
\end{equation}%
we have%
\begin{widetext}
\begin{align}
\sum_{l=0}^{\infty }\frac{\alpha ^{l}}{l!}H_{m+l,n+l}\left( x,y\right) &
=\sum_{l=0}^{\infty }\frac{\alpha ^{l}}{l!}(-1)^{n+l}e^{xy}\int \frac{d^{2}z%
}{\pi }z^{n+l}z^{\ast m+l}\exp \left[ -\left\vert z\right\vert
^{2}+xz-yz^{\ast }\right]   \notag \\
& =e^{xy}(-1)^{n}\int \frac{d^{2}z}{\pi }\sum_{l=0}^{\infty }\frac{\left(
-\alpha \left\vert z\right\vert ^{2}\right) ^{l}}{l!}z^{n}z^{\ast m}\exp %
\left[ -\left\vert z\right\vert ^{2}+xz-yz^{\ast }\right]   \notag \\
& =e^{xy}(-1)^{n}\int \frac{d^{2}z}{\pi }z^{n}z^{\ast m}\exp \left[ -\left(
\alpha +1\right) \left\vert z\right\vert ^{2}+xz-yz^{\ast }\right]   \notag
\\
& =\frac{e^{\frac{\alpha \allowbreak xy}{\alpha +1}}}{\left( \alpha
+1\right) ^{(m+n+2)/2}}(-1)^{n}e^{xy/\left( \alpha +1\right) }\int \frac{%
d^{2}z}{\pi }z^{n}z^{\ast m}\exp \left[ -\left\vert z\right\vert ^{2}+\frac{%
xz-yz^{\ast }}{\sqrt{\alpha +1}}\right]   \notag \\
& =\frac{e^{\frac{\alpha \allowbreak xy}{\alpha +1}}}{\left( \alpha
+1\right) ^{(m+n+2)/2}}H_{m,n}\left( \frac{x}{\sqrt{\alpha +1}},\frac{y}{%
\sqrt{\alpha +1}}\right) ,  \tag{D2}
\end{align}
\end{widetext}thus we have completed the proof of (C4).

\textbf{Appendix E: Derivation of Eq.(\ref{3.14})}

When $\chi =0,$ $\Lambda _{m,n}\rightarrow (1-e^{-2\gamma t})=T,$ and
\begin{align}
& C_{m,n}\left( \alpha ,\alpha ^{\ast },t\right)  \notag \\
& \equiv \frac{\exp \left[ -\gamma t\left( m+n\right) \right] }{m!n!\left(
T+1\right) ^{(m+n+2)/2}}H_{m,n}\left( 2\alpha ^{\ast },2\alpha \right)
e^{-2\left\vert \alpha \right\vert ^{2}},  \tag{E1}
\end{align}%
then we have
\begin{align}
& \sum_{m,n=0}^{\infty }C_{m,n}\left( \alpha ,\alpha ^{\ast },t\right)
H_{m,n}\left( 2\beta ,2\beta ^{\ast }\right)  \notag \\
& =\frac{e^{-2\left\vert \alpha \right\vert ^{2}}}{T+1}\sum_{m,n=0}^{\infty }%
\frac{\left( \frac{e^{-\gamma t}}{\sqrt{T+1}}\right) ^{m+n}}{m!n!}  \notag \\
& \times H_{m,n}\left( 2\alpha ^{\ast },2\alpha \right) H_{m,n}\left( 2\beta
,2\beta ^{\ast }\right) .  \tag{E2}
\end{align}%
Using the following formula%
\begin{align}
& \sum_{m,n=0}^{\infty }\frac{s^{m}t^{n}}{m!n!}H_{m,n}\left( x,y\right)
H_{m,n}\left( \alpha ,\beta \right)  \notag \\
& =\frac{1}{1-st}\exp \left[ \allowbreak \frac{sx\alpha +ty\beta -\left(
xy+\alpha \beta \right) st}{1-st}\right] ,  \tag{E3}
\end{align}%
Eq.(E2) can be rewritten as
\begin{equation}
\text{(E2)}=\frac{e^{-2\left\vert \alpha \right\vert ^{2}}}{2T}e^{\frac{%
\frac{4e^{-\gamma t}}{\sqrt{T+1}}\left( \alpha ^{\ast }\beta +\alpha \beta
^{\ast }\right) -\left( \alpha ^{\ast }\alpha +\beta \beta ^{\ast }\right)
\frac{4e^{-2\gamma t}}{T+1}}{1-\frac{e^{-2\gamma t}}{T+1}}}.  \tag{E4}
\end{equation}%
Thus Eq.(\ref{3.11}) becomes%
\begin{widetext}
\begin{align}
W\left( \alpha,\alpha^{\ast},t\right) & =4\int\frac{\mathtt{d}^{2}\beta }{\pi%
}e^{2\frac{T-1}{T+1}\left\vert \beta\right\vert
^{2}}\sum_{m,n=0}^{\infty}C_{m,n}\left( \alpha,\alpha^{\ast},t\right)
H_{m,n}\left( \frac{2\beta}{\sqrt{T+1}},\frac{2\beta^{\ast}}{\sqrt{T+1}}%
\right) W\left( \beta,\beta^{\ast},0\right)  \notag \\
& =\frac{2}{T}e^{-2\left\vert \alpha\right\vert ^{2}}\int\frac{\mathtt{d}%
^{2}\beta}{\pi}\exp\left[ 2\frac{T-1}{T+1}\left\vert \beta\right\vert ^{2}+%
\frac{2e^{-\gamma t}\left( \alpha^{\ast}\beta+\alpha\beta^{\ast}\right)
-2e^{-2\gamma t}\left( \left\vert \alpha\right\vert ^{2}+\frac{\left\vert
\beta\right\vert ^{2}}{T+1}\right) }{T}\right] W\left( \beta,\beta^{\ast
},0\right)  \notag \\
& =R.H.S.\text{ of Eq.(\ref{3.14}).}  \tag{E5}
\end{align}
\end{widetext}\textbf{Appendix F: Derivation of Eq.(\ref{5.4})}

Using the relation between Hermite polynomial and Lagurre polynomial,%
\begin{equation}
L_{m}\left( xy\right) =\frac{(-1)^{m}}{m!}H_{m,m}\left( x,y\right) ,
\tag{F1}
\end{equation}
we can recast the left of Eq.(\textbf{\ref{5.4}}) into the following form%
\begin{align}
F_{m,n} & =\frac{1}{s!}\int\frac{\mathtt{d}^{2}\alpha}{\pi}e^{-4\left\vert
\alpha\right\vert ^{2}}(-1)^{s}s!L_{s}\left( 4\left\vert \alpha\right\vert
^{2}\right) H_{m,n}\left( 2\alpha^{\ast},2\alpha\right)  \notag \\
& =\frac{1}{s!}\int\frac{\mathtt{d}^{2}\alpha}{\pi}e^{-4\left\vert
\alpha\right\vert ^{2}}H_{s,s}\left( 2\alpha^{\ast},2\alpha\right)
H_{m,n}\left( 2\alpha^{\ast},2\alpha\right)  \notag \\
& =\frac{1/4}{s!}\int\frac{\mathtt{d}^{2}\alpha}{\pi}e^{-\left\vert
\alpha\right\vert ^{2}}H_{s,s}\left( \alpha^{\ast},\alpha\right)
H_{m,n}\left( \alpha^{\ast},\alpha\right) .  \tag{F2}
\end{align}
Further using the generating function of $H_{m,n}\left( \epsilon
,\varepsilon\right) $,
\begin{equation}
H_{m,n}\left( \epsilon,\varepsilon\right) =\frac{\partial^{m+n}}{\partial
t^{m}\partial t^{\prime n}}\left. \exp\left[ -tt^{\prime}+\epsilon
t+\varepsilon t^{\prime}\right] \right\vert _{t=t^{\prime}=0},  \tag{F3}
\end{equation}
and the integration formula,%
\begin{equation}
\int\frac{d^{2}z}{\pi}e^{\zeta\left\vert z\right\vert ^{2}+\xi z+\eta
z^{\ast }}=-\frac{1}{\zeta}e^{-\frac{\xi\eta}{\zeta}},\text{Re}\left(
\zeta\right) <0,  \tag{F4}
\end{equation}
we have
\begin{align}
& \int\frac{\mathtt{d}^{2}\alpha}{\pi}e^{-\left\vert \alpha\right\vert
^{2}}H_{m^{\prime},n^{\prime}}\left( \alpha^{\ast},\alpha\right)
H_{m,n}\left( \alpha^{\ast},\alpha\right)  \notag \\
& =\frac{\partial^{m+n}}{\partial t^{m}\partial t^{\prime n}}\frac {%
\partial^{m+n}}{\partial\tau^{m}\partial\tau^{\prime n}}\exp\left[
-tt^{\prime}-\tau\tau^{\prime}\right]  \notag \\
& \times\int\frac{\mathtt{d}^{2}\alpha}{\pi}\left. \exp\left[ -\left\vert
\alpha\right\vert ^{2}+\left( t+\tau\right) \alpha^{\ast}+\left( t^{\prime
}+\tau^{\prime}\right) \alpha\right] \right\vert _{t=t^{\prime}=\tau
=\tau^{\prime}=0}  \notag \\
& =\frac{\partial^{m^{\prime}+n^{\prime}}}{\partial t^{m^{\prime}}\partial
t^{\prime n^{\prime}}}\frac{\partial^{m+n}}{\partial\tau^{m}\partial
\tau^{\prime n}}\exp\left[ t\tau^{\prime}+\tau t^{\prime}\right]
_{t=t^{\prime}=\tau=\tau^{\prime}=0}  \notag \\
& =m!n!\delta_{m^{\prime},n}\delta_{n^{\prime},m},  \tag{F5}
\end{align}
thus Eq.(F2) becomes the right hand side of Eq.(\textbf{\ref{5.4}}).\bigskip

\textbf{Appendix G: Derivation of Eq.(\ref{5.6})}

For this purpose, from Eq.(\ref{3.13}) we have
\begin{equation}
E_{m,n}=4\int \frac{\mathtt{d}^{2}\beta }{\pi ^{2}}e^{-2y\left\vert \beta
\right\vert ^{2}}H_{m,n}\left( 2x\beta ,2x\beta ^{\ast }\right)
e^{-2\left\vert \beta -z\right\vert ^{2}},  \tag{G1}
\end{equation}%
where we have set
\begin{equation}
y=\frac{1-\Lambda _{m,n}}{1+\Lambda _{m,n}},x=\frac{1}{\sqrt{\Lambda _{m,n}+1%
}}.  \tag{G2}
\end{equation}%
Using Eqs.(F3) and (F4), Eq.(G1) can be recast into the following form,%
\begin{widetext}
\begin{align}
E_{m,n} & =4e^{-2\left\vert z\right\vert ^{2}}\frac{\partial^{m+n}}{\partial
t^{m}\partial t^{\prime n}}e^{-tt^{\prime}}\int\frac{\mathtt{d}^{2}\beta}{%
\pi^{2}}\left. \exp\left[ -2\left( y+1\right) \left\vert \beta\right\vert
^{2}+2\beta\left( xt+z^{\ast}\right) +2\beta^{\ast}\left( xt^{\prime
}+z\right) \right] \right\vert _{t=t^{\prime}=0}  \notag \\
& =e^{-2\left\vert z\right\vert ^{2}}\frac{\partial^{m+n}}{\partial
t^{m}\partial t^{\prime n}}e^{-tt^{\prime}}\int\frac{\mathtt{d}^{2}\beta}{%
\pi^{2}}\left. \exp\left[ -\frac{y+1}{2}\left\vert \beta\right\vert
^{2}+\beta\left( xt+z^{\ast}\right) +\beta^{\ast}\left( xt^{\prime
}+z\right) \right] \right\vert _{t=t^{\prime}=0}  \notag \\
& =\frac{1}{\pi}\frac{2}{y+1}e^{-\frac{2y}{y+1}\left\vert z\right\vert ^{2}}%
\frac{\partial^{m+n}}{\partial t^{m}\partial t^{\prime n}}\exp\left[ -\left(
1-\frac{2x^{2}}{y+1}\right) tt^{\prime}+\frac{2xz}{y+1}t+\frac{2xz^{\ast}}{%
y+1}t^{\prime}\right] _{t=t^{\prime}=0}.  \tag{G3}
\end{align}
\end{widetext}Note that $1-\frac{2x^{2}}{y+1}=\allowbreak 0,y+1=\frac{%
1-\Lambda _{m,n}}{1+\Lambda _{m,n}}+1=\allowbreak \frac{2}{\Lambda _{m,n}+1}%
,-\frac{2y}{y+1}=\allowbreak \Lambda _{m,n}-1,$ then we find
\begin{align}
E_{m,n}& =\frac{1}{\pi }\frac{2}{y+1}e^{-\frac{2y}{y+1}\left\vert
z\right\vert ^{2}}\left( \frac{2xz}{y+1}\right) ^{m}\left( \frac{2xz^{\ast }%
}{y+1}\right) ^{n}  \notag \\
& =\frac{1}{\pi }\left( \Lambda _{m,n}+1\right) ^{\frac{m+n+2}{2}}e^{\left(
\allowbreak \Lambda _{m,n}-1\right) \left\vert z\right\vert
^{2}}z^{m}z^{\ast n}.  \tag{G4}
\end{align}

\bigskip

\bigskip

\end{document}